%% file: ics23.tex
\documentclass[conference]{IEEEtran}

\IEEEoverridecommandlockouts

\usepackage{algorithmic}
\usepackage{graphicx}
\usepackage{textcomp}
\usepackage{xcolor}
\definecolor{LightCyan}{rgb}{0.88,1,1}
\usepackage{comment}
\usepackage{acronym}
\usepackage{multirow}
\usepackage{svg}
\usepackage{amsmath}
\usepackage{subfigure}

\newcounter{usevspace}
\def\BibTeX{{\rm B\kern-.05em{\sc i\kern-.025em b}\kern-.08em
    T\kern-.1667em\lower.7ex\hbox{E}\kern-.125emX}}
\begin{document}

\title{Efficient Intra-Rack Resource Disaggregation for HPC Using Co-Packaged DWDM Photonics} 

\author{
\IEEEauthorblockN{George Michelogiannakis\IEEEauthorrefmark{1},
Yehia Arafa\IEEEauthorrefmark{2},
Brandon Cook\IEEEauthorrefmark{1},
Liang Yuan Dai\IEEEauthorrefmark{3},
\\
Abdel-Hameed (Hameed) Badawy\IEEEauthorrefmark{2},
Madeleine Glick\IEEEauthorrefmark{3},
Yuyang Wang\IEEEauthorrefmark{3},
Keren Bergman\IEEEauthorrefmark{3}, and
John Shalf\IEEEauthorrefmark{1}}
\IEEEauthorblockA{\IEEEauthorrefmark{1}Lawrence Berkeley National Laboratory, 
Berkeley, CA, USA\\
Email: \{mihelog,bcook,jshalf\}@lbl.gov}
\IEEEauthorblockA{\IEEEauthorrefmark{2}New Mexico State University, Las Cruces, NM, USA\\
Email: \{yarafa,badawy\}@nmsu.edu}
\IEEEauthorblockA{\IEEEauthorrefmark{3}Columbia University, New York, NY, USA\\
Email: \{ld2719,msg144,yw3831,kb2028\}@columbia.edu}}

\maketitle

\acrodef{RL}{race logic}
\acrodef{NoC}{network on chip}
\acrodef{RSFQ}{rapid single flux quantum}
\acrodef{SFQ}{single flux quantum}
\acrodef{FA}{first arrival}
\acrodef{LA}{last arrival}
\acrodef{I}{inhibit}
\acrodef{IQR}{interquartile range}
\acrodef{JTL}{Josephson transmission line}
\acrodef{JJ}{Josephson junction}
\acrodef{SQUID}{superconducting quantum interference device}
\acrodef{M}{merger}
\acrodef{S}{splitter}
\acrodef{DFF}{D flip flop}
\acrodef{UR}{uniform random}
\acrodef{WC}{worst case}
\acrodef{TTA}{transport-triggered architecture}
\acrodef{FF}{flip flop}
\acrodef{TFF}{toggle flip flop}
\acrodef{NDRO}{non destructive read out}
\acrodefplural{NoC}[NoCs]{networks on chip}
\acrodef{SRNoC}{superconducting rotary NoC}
\acrodef{GPU}{graphical processing unit}
\acrodef{ML}{machine learning}
\acrodef{LDMS}{light-weight distributed metric service}
\acrodef{OS}{operating system}
\acrodef{NIC}{network interface controller}
\acrodef{MIG}{multi-instance GPU}
\acrodef{HPC}{high performance computing}
\acrodef{CPU}{central processing unit}
\acrodef{CDF}{cumulative distribution function}
\acrodef{FPGA}{field programmable gate array}
\acrodef{IPC}{instructions per cycle}
\acrodef{LLC}{last level cache}
\acrodef{MRR}{micro ring resonator}
\acrodef{AWG}{arrayed waveguide grating}
\acrodef{DWDM}{dense wavelength division multiplexing}
\acrodef{BER}{bit error rate}
\acrodef{FOM}{figure of merit}
\acrodefplural{FOM}[FOMs]{figures of merit}
\acrodef{MCM}{multi chip module}
\acrodef{DFB}{distributed feedback}
\acrodef{FIT}{failures in time}
\acrodef{FEC}{forward error correction}
\acrodef{SiP}{silicon in-package photonic}
\acrodef{AMR}{adaptive mesh refinement}
\acrodef{OCS}{optical circuit switch}
\acrodefplural{OCS}[OCSs]{optical circuit switches}
\acrodef{MZI}{mach zehnder interferometers}
\acrodef{AWGR}{arrayed waveguide grating router}
\acrodef{SDN}{software defined network}
\acrodef{OS}{operating system}
\acrodef{NVM}{non-volatile memory}
\acrodef{NLP}{natural language processing}
\acrodef{BERT}{bidirectional encoder representations from transformers}
\acrodef{DLRM}{deep learning recommendation model}
\acrodef{CGRA}{coarse grain reconfigurable array}
\acrodef{HDD}{hard disk drive}
\acrodef{MSHR}{miss status handling register}
\acrodef{PIC}{photonic integrated circuit}
\acrodef{TSV}{through silicon via}
\acrodef{HBM}{high bandwidth memory}
\acrodef{SEC-DED}{single-error-correct/double-error-detect}
\acrodef{FEC}{forward error correction}
\acrodef{WDM}{wavelength-division multiplexing}
\acrodef{CXL}{compute express link}
\acrodef{MEMS}{microelectromechanical systems}
\acrodef{MZI}{Mach-Zehnder interferometer}
\acrodef{MRR}{microring resonator}
\acrodef{P2MP}{point-to-multipoint}
\acrodef{WSS}{wavelength-selective switching}
\acrodef{OEO}{optical electrical optical}
\acrodef{WPE}{wall plug efficiency}
\acrodef{OOO}{out-of-order}
\acrodef{DC-switch}{delivery-coupling switch}

\input{0_abstract}

\input{1_introduction}

\input{2_related_work}

\input{3_optical_switches}

\input{4_control_logic}

\input{5_system}

\input{6_evaluation}

\input{7_discussion}

\input{8_conclusion}

\bibliographystyle{IEEEtran}
\bibliography{bibliography_full}

\end{document}

%% file: 0_abstract.tex
\begin{abstract}

The diversity of workload requirements and increasing hardware heterogeneity in emerging \ac{HPC} systems motivate resource disaggregation. Resource disaggregation allows compute and memory resources to be allocated individually as required to each workload. However, it is unclear how to efficiently realize this capability and cost-effectively meet the stringent bandwidth and latency requirements of \ac{HPC} applications. To that end, we describe how modern photonics can be co-designed with modern \ac{HPC} racks to implement flexible intra-rack resource disaggregation and fully meet the \ac{BER} and high escape bandwidth of all chip types in modern \ac{HPC} racks.
Our photonic-based disaggregated rack provides an average application speedup of 11\% (46\% maximum) for 25 \acs{CPU} and 61\% for 24 \acs{GPU} benchmarks compared to a similar system that instead uses modern electronic switches for disaggregation. Using observed resource usage from a production system, we estimate that an iso-performance intra-rack disaggregated \ac{HPC} system using photonics would require 4$\times$ fewer memory modules and 2$\times$ fewer \acsp{NIC} than a non-disaggregated baseline.

\end{abstract}

%% file: 1_introduction.tex
\acresetall 

\section{Introduction}
\label{section:introduction}

Leading \ac{HPC} systems are steadily embracing heterogeneity of compute and memory resources to preserve performance scaling and reduce system power~\cite{Hybrid_overview, Top500, HPC_accelerators_list}.
This trend is already apparent with the integration of \acsp{GPU}~\cite{HPC_GPUs,HPC_GPUs2, HPC_GPUs3}
and is expected to continue with fixed-function or reconfigurable accelerators such as \acp{FPGA}~\cite{AI_accelerator, Deep_learning_accelerator, FPGAs, FPGA_fluids, HPC_FPGAs3, FPGA_promise, ML_accelerator}, and heterogeneous memory~\cite{Heterogeneous_memory}.
Also, key \ac{HPC} workloads show considerable diversity in computational and memory access patterns~\cite{TACO_disaggregation, Job_heterogeneity}.

This expectation of resource heterogeneity, workload diversity, and today's method of allocating resources to applications in units of statically-configured nodes where every node is identical and unused resources are left to idle (referred to as ``marooned resources''), raises the concern of resource underutilization. Marooned resources increase capital and operational costs without improving performance. This has motivated resource disaggregation. Disaggregation refers to decomposing servers into their constituent compute and memory resources so that these can be allocated as required according to the needs of each workload. Hyperscale datacenters have embraced resource disaggregation and showed that it significantly improves utilization of \acsp{GPU} and memory~\cite{GPU_disaggregation, Memory_disaggregation, Facebook_rack_disaggregation, Azure_disaggregation, Elastic_memory, Datacenter_disaggregation, Optics_disaggregated_memory, Disaggregation_datacenters, GPU_rack_level,Disaggregation_pond,Disaggregation_page_policy,Disaggregation_tiered_memory}.

Although file storage is routinely disaggregated in modern systems~\cite{Perlmutter_configuration, TACO_disaggregation, HDD_disaggregation},
\ac{HPC} has been slow to embrace disaggregation of compute and memory resources~\cite{PINE, FOSDA} due to the sensitivity of \ac{HPC} workloads to bandwidth and latency that cannot be met by current PCIe/CXL or Ethernet link technologies used in contemporary disaggregated architectures. Studies showed that disaggregation only among resources in the same rack (i.e., intra-rack resource disaggregation) in \ac{HPC} could reduce resources by 5.36\% to 69.01\% while avoiding the overhead of full-system disaggregation~\cite{TACO_disaggregation}. However, the impact of increased memory latency and specific architectural trade-offs have not been explored. Thus, although disaggregation using electronic networks has been demonstrated in hyperscale datacenters~\cite{Datacenter_disaggregation2, Datacenter_disaggregation, Disaggregation_through_SDNs,Disaggregation_pond},
minimizing adverse effects to
and addressing the stringent bandwidth density and latency demands of
\ac{HPC} workloads requires a thorough investigation.

Our contributions are as follows. Firstly, we describe how to use emerging photonic links and switches to design modern and practical resource-disaggregated \ac{HPC} racks based on an existing \acs{GPU}-accelerated HPE/Cray EX supercomputer~\cite{Perlmutter_configuration}. Secondly, we show how state-of-the-art commercially available photonic components and advanced packaging \acp{MCM} meet \ac{BER} requirements, impose only a 5\% power overhead, and deliver sufficient bandwidth to satisfy the escape bandwidth of all chips in modern \ac{HPC} racks. Thirdly, we show how to use distributed indirect routing and \acp{AWGR}~\cite{Yoo20, AWGR_all_to_all} to satisfy all bandwidth requirements without the overhead and latency for reconfiguration that spatial~\cite{MingWuSwitch, Spatial7} and wave-selective~\cite{Hybrid_switch} switches require. Furthermore, we show that intra-rack disaggregation using emerging photonics provides an average application speedup of 11\% (46\% maximum) for 25 \acs{CPU} and 61\% for 24 \acs{GPU} benchmarks compared to a similar system that instead uses state-of-the-art electronic switches. 
Finally, based on observed resource usage, we estimate that a system based on state-of-the-art photonics for resource disaggregation can have 4$\times$ fewer memory modules and 2$\times$ fewer \acsp{NIC}, thus 44\% fewer overall chips compared to a non-disaggregated system with the same computational throughput.

\acresetall 

%% file: 2_related_work.tex
\section{Related Work}
\label{section:related work}

Hyperscale datacenters predominantly focus on full-system resource disaggregation where applications can allocate fine-grain resources of different types, today typically \acp{GPU}~\cite{GPU_disaggregation} and memory~\cite{Memory_disaggregation,2009_disaggregation, Elastic_memory, Optics_disaggregated_memory,Disaggregation_pond,Disaggregation_tiered_memory}.
In such a system, resources of the same type are typically placed in the same rack~\cite{Datacenter_disaggregation2, Datacenter_disaggregation, Disaggregation_through_SDNs}.

However, full-system, flexible, and fine-grain resource disaggregation introduces significant overhead because of the higher latency and lower bandwidth density of contemporary hardware used to implement resource disaggregation -- typically PCIe, 100Gig Ethernet, and eventually \ac{CXL}~\cite{CXL} over electronic links.
This overhead does not simply increase power and procurement costs. Instead, it adds potentially substantial latency between key resources such as \acp{CPU} and memory traditionally exhibit
latency-sensitive communication. The aforementioned studies quote several orders of magnitude increase in network and memory latency due to full-system resource disaggregation to improve resource utilization by 35\% at most~\cite{Disaggregation_optics}.
Another study found that application performance degradation depends on both network bandwidth and latency, but can still reach 40\% even with high bandwidth, low-latency networks~\cite{Disaggregation_network_requirements}.
Work on SPEC and commercial benchmarks also found an up to 27\% application slowdown due to the additional memory latency~\cite{Abali}. A study on Microsoft's Azure found a range of performance slowdowns up to 30\% from an extra 65 ns to access main memory~\cite{Azure_disaggregation}; a later study reported a range of slowdowns with a mean of 10\%, a higher average, and a maximum of about 100\% from an additional 142 ns. \Acp{SDN} based on electrical networks fare no better in terms of overhead~\cite{Disaggregation_network_requirements, Disaggregation_hotnets,  Disaggregation_through_SDNs}.

Hybrid full-system photonic--electronic approaches have also been proposed that rely on circuit switching~\cite{Disaggregation_optics}.
A few studies argue that intra-rack disaggregation~\cite{Facebook_rack_disaggregation, 2009_disaggregation, GPU_rack_level} or disaggregation among small groups of \acp{CPU}~\cite{Disaggregation_pond,Disaggregation_tiered_memory} suffice to recover most gains.
Even the low latency and high bandwidth density of modern photonics can only partially satisfy the bandwidth, energy, and latency requirements of full system disaggregation. This makes system-wide disaggregation impractical in many cases~\cite{Datacenter_disaggregation2,  Disaggregation_optics, Disaggregation_datacenters, Disaggregation_optical_limits}.

Recent full system approaches in \ac{HPC} rely on optics to connect \acp{CPU} and memory, and electronic switches for \acp{HDD} to increase resource \ac{CPU} utilization by 36.6\% and memory 21.5\%~\cite{FOSDA}.
Another study argues that \ac{HPC} systems can reduce resources from 5.36\% to 69.01\% with intra-rack disaggregation and still satisfy the worst-case average rack utilization~\cite{TACO_disaggregation}.
Similar to datacenters, intra-rack disaggregation in \ac{HPC} promises the lowest overhead and impact to applications~\cite{PINE, Facebook_rack_disaggregation, GPU_rack_level}.

Related work has researched other aspects necessary to make resource disaggregation practical in a system, such as job scheduling~\cite{Scheduling_heterogeneity, Heterogeneous_resource_allocation, Resource_allocation_disaggregation, Heterogeneous_scheduler1}, how the \ac{OS}
and runtime should adapt~\cite{Runtime_heterogeneity, Runtimes_for_heterogeneous, LegoOS}, page migration policies and temporal imbalance~\cite{Disaggregation_page_policy,Disaggregation_tiered_memory},
programming and code portability in heterogeneous systems~\cite{Programming_heterogeneity, Heterogeneous_resource_allocation},
partitioning of application data~\cite{Partitioning_applications},
fault tolerance~\cite{Reliability_heterogeneity},
how to fairly compare the performance of different heterogeneous systems~\cite{Benchmarking_heterogeneous},
and the impact of heterogeneous resources to application performance~\cite{Running_on_heterogeneous, Data_parallel_heterogeneity, Heterogeneous_memory}.
These are important but out of scope topics for our study.

\subsection{Under-utilization in Production Systems}
\label{section:underutilization in production systems}

We use NERSC's Cori as an exemplar production \ac{HPC} system due to its diverse and open-science workload, while recognizing workload requirements on other systems may differ.
In NERSC's Cori, at a time before Perlmutter became available and thus Cori was serving the full NERSC workload, three quarters of the time, Haswell nodes use less than 17.4\% of memory capacity (50.1\% for KNL nodes) and less than 0.46 GB/s of memory bandwidth~\cite{TACO_disaggregation}. These observations are similar to observations made on LANL clusters~\cite{Memory_disaggregation} and Alibaba machines that execute batch jobs. Likewise, half of the time, Cori nodes use no more than half of their compute cores and three quarters of the time 1.25\% of available \ac{NIC} bandwidth.
Similarly, in Lawrence Livermore National Laboratory clusters, approximately 75\% of the time, no more than 20\% of memory capacity is used~\cite{Memory_disaggregation}.
Alibaba's published data~\cite{Alibaba} show that memory is underutilized similar to Cori, for machines that execute batch jobs.
Data from Google systems shows that task memory and disk capacity is spread over three orders of magnitude and typically underutilized~\cite{Disaggregation_hotnets}.
Azure reports approximately 25\% of memory under-utilization~\cite{Azure_disaggregation,Disaggregation_pond}.
Datacenters have also reported 28\% to 55\% \ac{CPU} idle in the case of Google trace
data~\cite{CPU_cloud_utilization} and 20\%--50\% usually
in Alibaba~\cite{Alibaba}.
Early studies also suggest \ac{GPU} under-utilization~\cite{GPU_sharing_HPC2, GPU_profiling_2019, GPU_sharing_HPC}.

%% file: 3_optical_switches.tex
\section{Photonics for Resource Disaggregation}
\label{section:optical technologies for resource disaggregation}

Here we show that photonic links and switches today meet the strict performance and error rate requirements to efficiently implement intra-rack resource disaggregation in \ac{HPC}.

\subsection{Memory Technologies and Requirements}
\label{section:memorytechnology}

IO systems in \ac{HPC} are already largely disaggregated over conventional system-scale interconnects since the underlying technologies (disk or SSD) are relatively high latency and lower bandwidth~\cite{TACO_disaggregation, AWGR_pleros}.
In contrast, memory technologies, particularly \ac{HBM} needed by \acp{GPU}, are much higher bandwidth and much less tolerant of latency and require much lower \acp{BER}. 
Given that memory disaggregation imposes the most challenging constraints among other resources in today's compute nodes, we will use DDR and \ac{HBM} memory technology to set our performance target.
A typical DDR4 memory has a response latency of approximately 90 ns, and for \ac{HBM}, the average response latency is 90-140 ns~\cite{BenchHBM2020}. Still, any added latency between the \ac{CPU} and memory from resource disaggregation may penalize application performance, as we quantify later.
Server-class memories typically require \acp{BER} of less than $10^{-18}$ to achieve tolerable \ac{FIT} rates with conventional \ac{SEC-DED} protection~\cite{Facebook2015, GoodBadUgly2014}.
\Ac{FEC} can reduce the \ac{BER}, but with additional latency~\cite{FEC}.

\subsection{Optical Link Technologies}
\label{section:link technologies}

\begin{figure*}
  \centering
  \includegraphics[width=\textwidth]{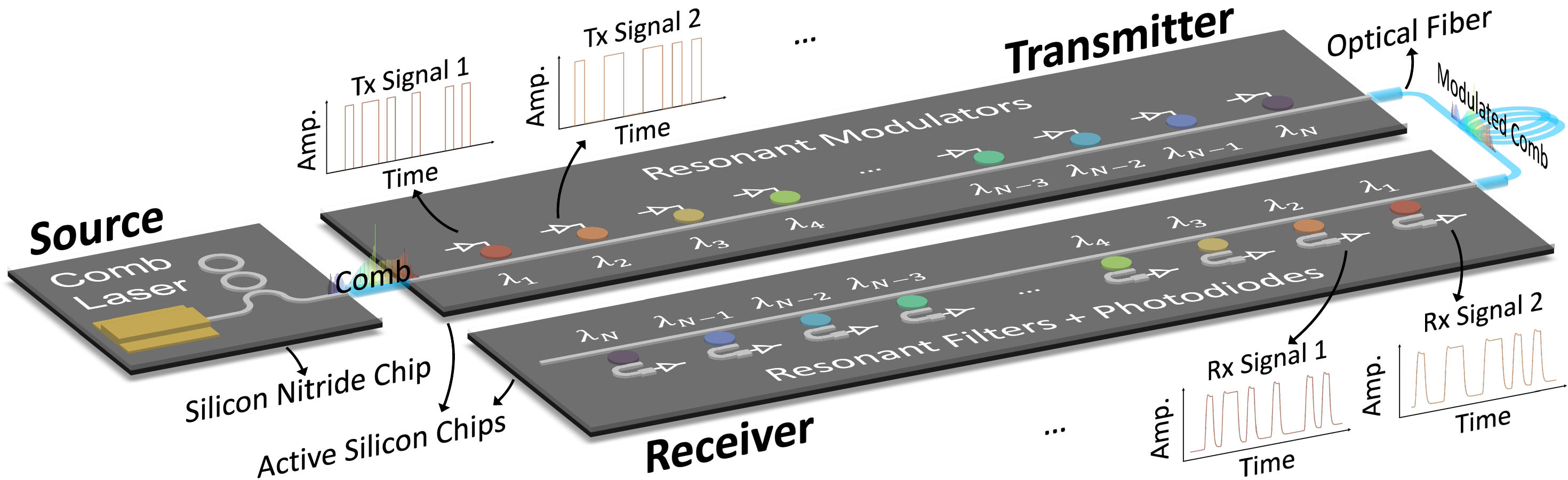}
  \ifnum\value{usevspace}>0
  \vspace{-0.4cm}
  \fi
  \caption{Logical schematic of a DWDM link using ring resonators and a comb-laser source. Each ring is tuned to a different frequency of light and can be used to modulate that specific wavelength of light (a channel). Comb laser sources provide a comb of frequencies of light to provide those wavelengths for encoding. All encoded optical channels share the same optical fiber and are decoded using the rings on the receiving side to route channels to the photodetectors.\label{figure:DWDMlink}}
  \ifnum\value{usevspace}>0
  \vspace{-0.6cm}
  \fi
\end{figure*}

We consider a range of photonic link technologies
that include conventional 100 Gbps Ethernet physical interfaces that represent the current baseline link technology for memory disaggregation. We also
introduce a range of cutting-edge \ac{DWDM} link technologies that are either demonstrated as research prototypes or are commercially available.
All photonic components come from existing commercial technologies (100 Gbps, 400 Gbps, Ayar TeraPhy) and some research prototypes from DARPA PIPES (the 1-2 Tbps link technologies).
These higher performance link technologies must be co-packaged to achieve their bandwidth density. These link technologies are summarized in Table~\ref{table:photonic link summary}.
The technology for the optical links is depicted in \figurename~\ref{figure:DWDMlink}.
Delivering multiple channels of laser light to the package has been challenging to scale cost-effectively if each ``color'' of light were to require a separate laser source. This concern was alleviated by the emergence of quantum dot and soliton comb laser sources
that can produce hundreds of usable light frequencies with wall-plug efficiencies of up to 41\%~\cite{Gaeta}.

\begin{figure*}
  \begin{center}
    \subfigure[Active optical MCM]{
      \includegraphics[width=0.31\textwidth]{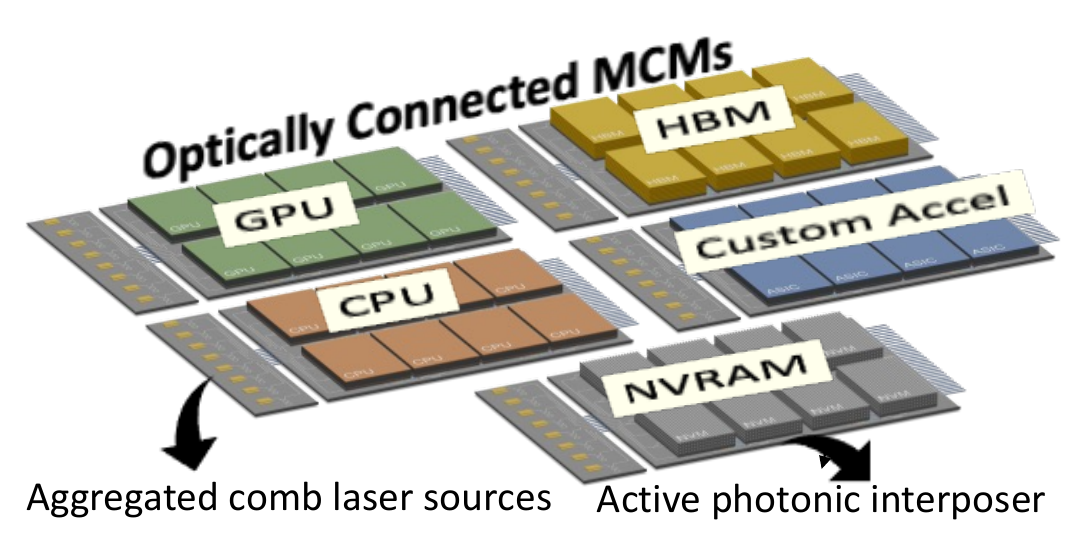}
    }
    \subfigure[Blade]{
      \includegraphics[width=0.31\textwidth]{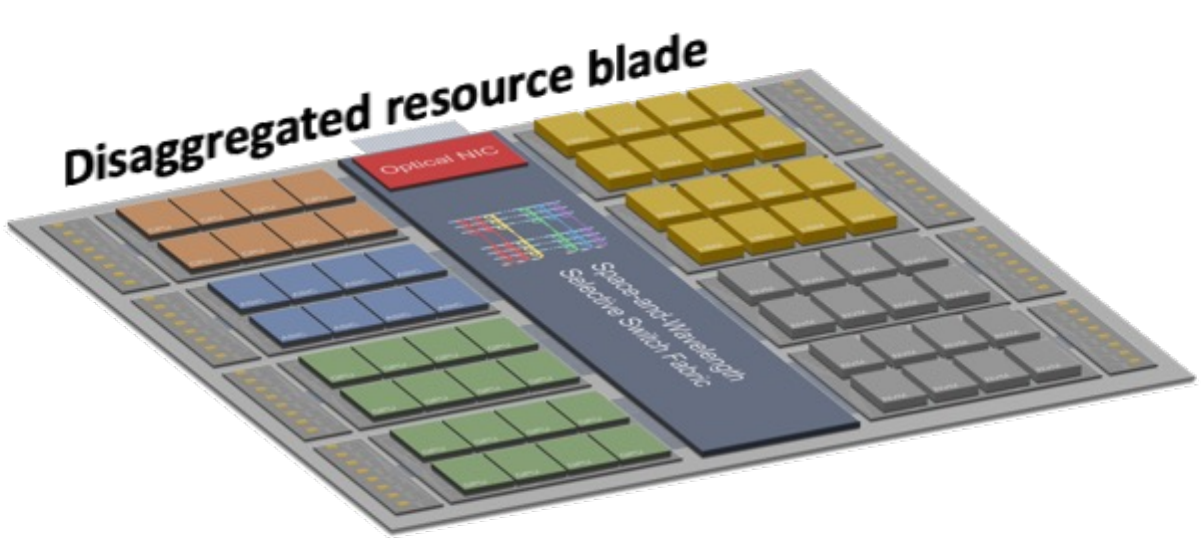}
    }
    \subfigure[Rack/Pod]{
      \includegraphics[width=0.27\textwidth]{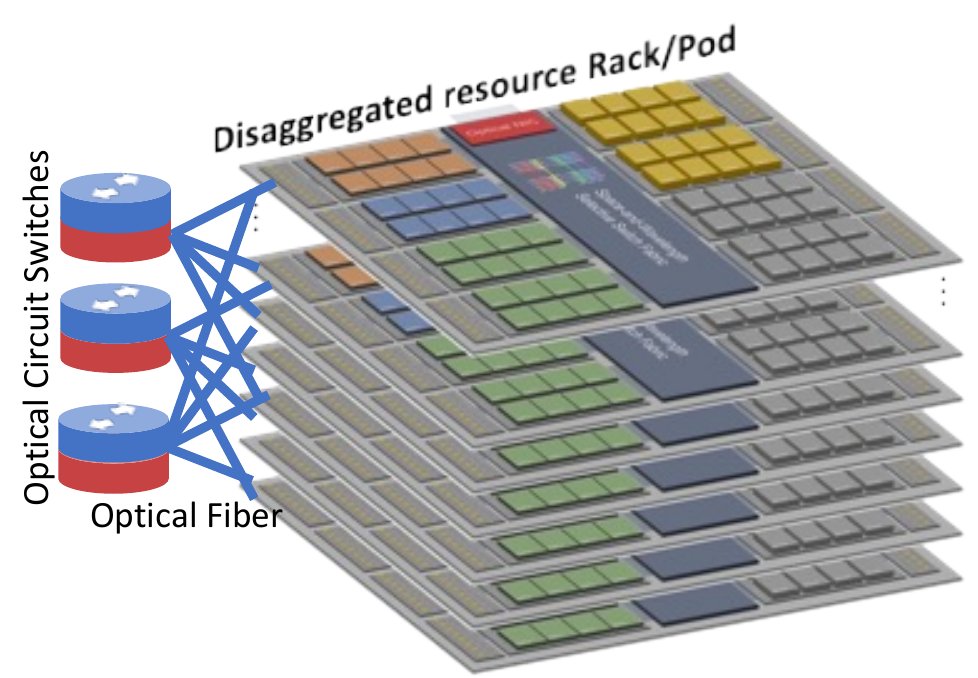}
    }
    \ifnum\value{usevspace}>0
    \vspace{-0.4cm}
    \fi
    \caption{Overall physical structure of rack (also referred to as pod) scale resource disaggregation from photonically-connected \acsp{MCM} up to the entire rack scale. The conversion from \acs{CXL}-over-fiber to \ac{HBM} or \ac{NVM} electrical protocol is implemented in the active interposer for the photonics \acs{MCM}.\label{figure:OpticalMCM}}
    \ifnum\value{usevspace}>0
    \vspace{-0.8cm}
    \fi
  \end{center}
\end{figure*}

\begin{figure}
\centering
  \includegraphics[width=\columnwidth]{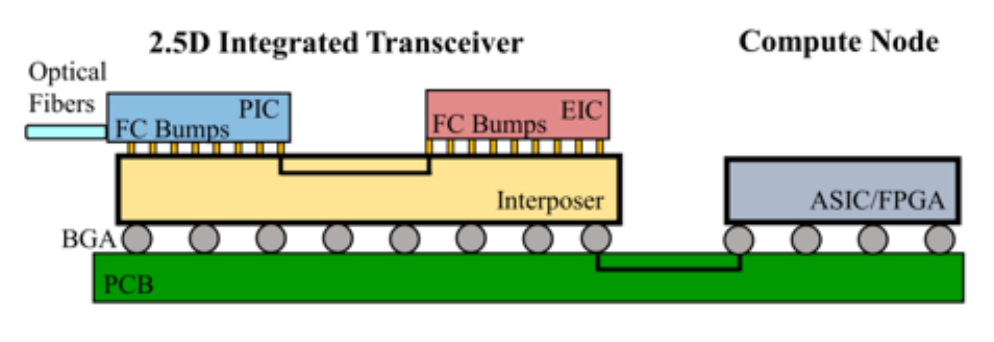}
  \ifnum\value{usevspace}>0
  \vspace{-1.0cm}
  \fi
  \caption{Co-packaged optics are required for \acs{DWDM} link technologies to achieve the bandwidth density to operate at native memory bandwidths.\label{figure:CopackagedOptics}}
  \ifnum\value{usevspace}>0
  \vspace{-0.4cm}
  \fi
\end{figure}

\subsection{Active Photonic MCMs}

Many \acp{CPU} and \acp{GPU} do not have the necessary off-chip bandwidth for full utilization of their compute resources because operating their I/O pins at a higher bandwidth incurs a power cost~\cite{Pin_limitation, TPU}.
Using emerging high-speed optical links directly to the \ac{MCM}, illustrated in \figurename~\ref{figure:CopackagedOptics}, provides to the order of $10\times$ gains in escape bandwidth~\cite{PINE, TeraPHY, PINE_OI, Co_packaged_fibers}.
This is a necessary property to enable efficient resource disaggregation as well as handle changing bandwidth requirements of key applications such as machine learning that drastically shifts bandwidth between inter-\acp{GPU} and off-chip from inference to training.

\acp{MCM} with integrated photonics have been demonstrated in both 2.5D and 3D interposer platforms~\cite{PINE,MCM_photonic,MCM_photonic3,MCM_photonic2}.
They can use different die-to-die link standards, such as UCIe. Active interposer platforms combine the \ac{PIC} and interposer into a single integrated substrate. The active interposer allows photonic components to be fabricated and directly integrated with \acp{TSV} and additional metal redistribution layers. Electronic circuits are flip-chipped on top of active interposers using copper pillars~\cite{Copper_pillars}. Further work has embedded photonic switch fabrics within \ac{MCM} platforms with a crosstalk suppression and extinction ratio of $>$50dB and on-chip loss as low $<$1.8dB~\cite{PINE}. This was further scaled up to support more than 100 ports with microring resonators using a scalable switch fabric that combined switching in the space domain with wavelength-selectivity to implement fine-grain connectivity for node disaggregation~\cite{Hybrid_switch, PINE}.

\begin{table}
  \begin{center}
    \begin{tabular}{|p{0.75cm}|p{0.75cm}|p{1.25cm}|p{1cm}|p{1cm}|p{1cm}|p{1cm}|}
      \hline
      BW (Gbps) & Energy (pJ/bit) & Link Gbps $\times$ Channels & \#Links (2 TB/s escape) & Agg. Ws (2 TB/s escape) & Ref.                                    \\
      \hline
      \hline
      100       & 30              & $25 \times 4$               & 160                     & 480                     & \cite{Commercial_photonics,Agrell_2016} \\ 
      \hline
      400       & 30              & $100 \times 4$              & 40                      & 197                     & \cite{OptCom400Gig}                     \\ 
      \hline
      768       & $<$ 1           & $32 \times 24$             & 21                      & 14.4                    & \cite{TeraPHY}                          \\ 
      \hline
      1,024     & 0.45            & $16 \times 64$              & 16                      & 7.2                     & \cite{Kim19}                            \\ 
      \hline
      2,048     & 0.3             & $16 \times 128$             & 8                       & 4.8                     & \cite{Kim19}                            \\ 
      \hline
    \end{tabular}
  \end{center}
  \caption{A range of \acs{WDM} photonic link technologies.\label{table:photonic link summary}}
  \ifnum\value{usevspace}>0
  \vspace{-1.0cm}
  \fi
\end{table}

\subsubsection{Link Protocol}

We adopt \ac{CXL} as our link protocol~\cite{CXL}. \ac{CXL} is an overlay on the PCIe-Gen6 physical layer; it includes guaranteed ordering of events and is a broadly adopted industry standard with published specifications.
However, we do not rely on any features of any particular protocol. Thus, alternatives such as UCIe also apply.

\subsubsection{Link Propagation and Encoding/Decoding Latency}
\label{section:link propagation and encoding/decoding latency}

The target reach for an intra-rack disaggregation solution is approximately 1-4 meters. Given the speed of light $c$ and light propagating through optical material with an index of refraction near r1.5, the effective latency of propagating through an optical fiber at nominally 0.75$c$ is approximately 5 ns per meter. Therefore, rack-scale resource disaggregation adds 5-20 ns of latency, approximately less than 20\% of the typical DRAM latency. The link latency for SERDES and photonic ring modulation is negligible.
Intra-rack fiber lengths up to 4 meters require no intervening \ac{OEO} conversions.

\subsubsection{Bit Error Rates and FEC}
\label{section:bit error rates and fec}

To achieve $10^{-18}$ \ac{BER} required for memory technologies, \ac{FEC}~\cite{FEC} will likely be required. Using the lightweight \ac{FEC} scheme that is proposed for \ac{CXL}~\cite{CXL} and PCIe Gen6~\cite{FEC_PCIe6} as an example, the all-inclusive latency for \ac{FEC} can be as low as 2 ns.  
Therefore, for 200 Gbps, the serialization delay is 10 ns and the \ac{FEC} calculations add 2-3 ns. At 400 Gbps and above, the net latency for \ac{FEC} would be 5 ns plus 2-3 ns. Notably, this approach to achieving these \ac{BER} targets is achievable with less than a 0.1\% bandwidth loss.

In terms of impact on \ac{BER}, this PCIe/CXL-like correction scheme corrects all single bursts of up to 16 bits. Double bursts will likely be mis-corrected, but the chance of a bad flit decreases quadratically
(e.g., a flit \ac{BER} of $10^{-6}$ becomes $10^{-12}$ as you need two error bursts per flit to fail).
Each flit is protected with a strong 64-flit CRC such that the flit \ac{FIT} rate (CRC escapes) is significantly less than one part per billion. Lastly, \ac{FEC} escapes become link retransmissions and the ASIC-to-ASIC connection sees close to zero errors. As a result, emerging memory fabric protocols such as \ac{CXL}, which could be run over our evaluated physical links, \emph{are capable} of achieving a \ac{BER} rate that meets the stringent memory system requirements and minimizes performance loss due to retransmission.

\subsection{Optical Switch Technologies}
\label{section:switch technologies}

Motivated by minimizing latency, our vision for a disaggregated rack is to have photonically-enabled \acp{MCM} that are connected via an optical circuit switch, as shown in \figurename~\ref{figure:OpticalMCM}. Compute and memory chips would be in the center of the \ac{MCM} and the edge of the \ac{MCM} would contain co-packaged optical \acp{SiP}.
Switches with all-optical paths include spatial- and wave-selective approaches, shown in Table~\ref{table:photonic switch summary}.

\begin{table}
  \begin{center}
    \begin{tabular}{|p{1.4cm}|p{1.1cm}|p{0.8cm}|p{1.1cm}|p{0.9cm}|p{1cm}|}
      \hline 
      Switch Type                                                         & Radix                       & Wave-lengths per port & B/W per channel (wavelength) & Insertion Loss & Crosstalk \\
      \hline
      \hline
      Mach-Zehnder based~\cite{IkedaMZI-2020}                             & $32 \times 32$                & 1                     & 439 Gbps                     & 12.8 dB        & -26.6 dB  \\ 
      \hline
      MEMS-actuated~\cite{240x240WaferScale}                              & $240 \times 240$              & 1                     & --                           & 9.8 dB         & -70 dB    \\ 
      \hline
      Microring resonator~\cite{ElasticWDMxBar2017,ScalableMicroringClos} & $8 \times 8$ ($128 \times 128$) & 8 (128)               & 100 Gbps (42 Gbps)           & 5dB (10dB)     & (-35 dB)  \\ 
      \hline
      Casc. AWGRs~\cite{satoRealizationApplicationLargeScale2018}         & $370 \times 370$              & 370                   & 25 Gbps                      & ~15 dB         & -35 dB    \\ 
      \hline
    \end{tabular}
  \end{center}
  \caption{High-radix CMOS-compatible photonic switches.
    \label{table:photonic switch summary}}
  \ifnum\value{usevspace}>0
  \vspace{-1.0cm}
  \fi
\end{table}

\subsubsection{Spatial Optical Switches}
\label{section:spatial optical switches}

In recent years, the primary switching cells investigated are \ac{MEMS} actuated couplers, \acp{MZI}, and \acp{MRR}.
Taking after their free-space counterpart, photonic \ac{MEMS}-actuated switches are broadband spatial switches that have demonstrated radix scaling up to $240 \times 240$~\cite{240x240WaferScale}. Although they typically offer low inter-channel crosstalk and low optical losses, \ac{MEMS} switching cells generally require high driving voltages (greater than 20 V), making them less attractive for co-integration with electronic drivers. Spatial switches can also use mirrors~\cite{Calient}, photonic integrated circuits~\cite{Spatial7}, or tiled planar silicon photonics~\cite{MingWuSwitch}.
\ac{MZI} switches are more friendly to co-integration compared to \ac{MEMS} but have only been shown to scale up to $32 \times 32$~\cite{IkedaMZI-2020}.
This limit can be seen as a consequence of the higher insertion-loss scaling resulting from cascaded \ac{MZI} cells and the susceptibility of popular \ac{MZI} topologies to first-order crosstalk.

The challenge for scaling up the spatial approach is the quantization of package and \ac{MCM} escape bandwidth and reduced configuration options. For example, at 768 Gbps (the Ayar TeraPhy~\cite{TeraPHY}), the number of fibers escaping the package is 21, meaning the package can be connected only up to 21 different potential destinations using a spatial switch.

\subsubsection{Wavelength Selective Optical Switches and AWGRs}
\label{section:wavelength selective optical switches and awgrs}

The inherent wavelength-selectivity of \ac{MRR} switching cells allows for the straightforward implementation of \ac{WSS} topologies. This enables one to establish all-to-all networks by leveraging \ac{WDM}. Currently, \ac{MRR}-based switches with the largest radix include the $8 \times 8$ crossbar~\cite{ElasticWDMxBar2017} and switch-and-select~\cite{NonBlockingSwitchFabric}, but have been experimentally emulated to include a $16 \times 16$ Clos~\cite{PAM4CLOS}. The metrics in~\cite{PAM4CLOS} can be seen to correlate very closely with the scaling proposed in~\cite{ScalableMicroringClos}, making a practical case for the $128 \times 128$ shown in Table~\ref{table:photonic switch summary}.

All-to-all networks via \ac{WDM} signals can also be achieved by \acp{AWGR}~\cite{Yoo20,AWGR_all_to_all,AWGR_all_to_all2, AWGR, AWGR_pleros}.
As \acp{AWGR} are passive optical elements, no reconfiguration is possible within the routing fabric itself. Instead, fast wavelength-tunable lasers must be leveraged at the transmitter of every node if it wishes to address a different destination since
\acp{AWGR} shuffle the light frequencies such that one lambda goes to each endpoint from each source.
\acp{AWGR} enable us to implement an $N \times N$ all-to-all topology using just $\mathcal{O}(N)$ fibers (each carrying $N$ frequencies of light). In contrast, an implementation using copper would require $N^2$ wires.
Although the cost of fast wavelength-tunable lasers is still an ongoing research topic~\cite{TunableLasers2019},
\acp{AWGR} are mature, commercially available, and well established in literature~\cite{FSproduct}.

In \acp{AWGR}, only a limited number of ports can be practically supported due to the walk-off of passband center frequencies from the carrier wavelength grid and the worse crosstalk associated with a larger number of ports ($N$). A feasible implementation of  \acs{AWGR}-based optical switches with a large $N$ has been demonstrated utilizing cascaded small-size \acp{AWGR}~\cite{satoRealizationApplicationLargeScale2018}. Specifically, $N$ $M \times M$ \acp{AWGR} (front-\acp{AWGR}) are interconnected with $M$ $N \times N$ \acp{AWGR} (rear-\acp{AWGR}) to effectively act as an $MN \times MN$ \ac{AWGR}. Each output port of a front-\ac{AWGR} is connected to an input port of a rear-\ac{AWGR}, where the interconnection pattern can be optimized with knowledge of port-specific insertion losses to minimize the worst-case end-to-end insertion loss. Further up-scaling of the switch radix can be achieved by interconnecting small $K \times K$ \acp{DC-switch} with multiple copies of the $MN \times MN$ \acp{AWGR}, yielding a $KMN \times KMN$ switching capability. This architecture has been verified by hardware prototypes of $270 \times 270$ and $1440 \times 1440$~\cite{satoLargescaleWavelengthRouting2013,uedaDemonstration4404402016}, showing $\sim$15\,dB insertion loss and below $-35$\,dB crosstalk suppression. In order to accommodate the 350 \acp{MCM} of our rack, a reasonable configuration is $KMN=3 \times 12 \times 11 = 396$. This results in 370 ports and 370 wavelengths per port (Table~\ref{table:photonic switch summary}). Since \acp{AWGR} typically have a 25 GHz optical bandwidth if the wavelength grid is 50 GHz, with PAM4, we assume 25 Gbps per wavelength~\cite{PAM4CLOS, PAM4}.

Wave-selective switches~\cite{Hybrid_switch,Photonic_survey}
can steer any subset of wavelengths to a given destination, not just all (spatial) or one (\ac{AWGR}).
Dynamic programming methods can avoid sending the same frequency of light from two different sources to the same destination.
Since this is a relatively new technology,
we constructed a model shown in Table~\ref{table:photonic switch summary} that projects the performance of a larger radix switch comprised of smaller demonstrated building blocks.

\subsubsection{Reconfiguration Time}
\label{section:reconfiguration time}

Spatial and wave-selective switches typically require centralized scheduling~\cite{TAGO} to reach a steady globally optimal solution.
The reconfiguration time can range from tens of nanoseconds to tens of milliseconds. In production \ac{HPC} systems, multi-node jobs start every few seconds and last from minutes to hours~\cite{Bandwidth_steering, TACO_disaggregation}. Also, job resource usage and communication become predictable early, do not change fast, and typically remain predictable throughout a job's execution time~\cite{TACO_disaggregation, Bandwidth_steering, ShalfKOS05, VetterM02}.
Therefore, even milliseconds of reconfiguration time is ample.

%% file: 4_control_logic.tex
\section{Control Logic}
\label{section:control logic}

Here we describe how we can perform indirect routing to increase point-to-point bandwidth using only per-source logic.

\begin{figure}
  \centering
  \includegraphics[width=0.8\columnwidth]{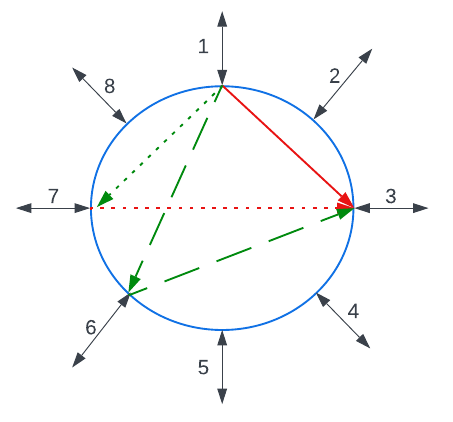}
  \ifnum\value{usevspace}>0
  \vspace{-0.6cm}
  \fi
  \caption{With an \ac{AWGR}, source 1 has one wavelength directly connecting it to source 3. For more bandwidth, it can route through another intermediate source (indirect routing) chosen in a Valiant fashion~\cite{Yoo20, TAGO}. Here, the link from 1 to 7 is available (green), but the link from 7 to 3 is not (red). The chosen path is from 1 to 6 to 3 because both links are available.\label{figure:awgr routing}}
  \ifnum\value{usevspace}>0
  \vspace{-0.6cm}
  \fi
\end{figure}

\subsection{Indirect routing in AWGRs}
\label{section:indirect routing in AWGRs}

\acp{AWGR} dedicate exactly one wavelength between any source--destination pair. If a source--destination pair requires more bandwidth than what a single wavelength can satisfy, sources can use indirect routing, an example of which is shown in \figurename~\ref{figure:awgr routing}.
Sources can split traffic to $N$ intermediate destinations in parallel in order to use the bandwidth of $N$ wavelengths.
This does not consume additional power in the photonic components assuming lasers are constantly powered.
Sources consider indirect paths only if the direct (single-hop) bandwidth to their desired destination does not suffice. A source considers indirect destinations for which the direct bandwidth from the source is available and whose wavelength from the intermediate hop to the desired final destination is also available. Among potentially multiple candidates, sources can choose one in a Valiant fashion~\cite{Yoo20, TAGO, HyperX_demonstrated}.
This is done on a per-flow basis in order to avoid out of order packet delivery.
This routing can be modeled as a well-established allocation problem and implemented with a low latency and area penalty~\cite{Routing_logic_implementation, Allocator_implementations}.

Indirect routing relies on sources knowing which other sources attached to the same \ac{AWGR} are utilizing their local wavelengths in order to identify a productive intermediate destination.
For instance, in \figurename~\ref{figure:awgr routing}, source 1 should know whether the wavelengths from 7 to 3 and 6 to 3 are occupied.
For that, we rely on piggybacking, where traffic between a source--destination pair periodically includes the state of the sources' wavelengths as a way to broadcast the local state to the rest of the sources attached to the same \ac{AWGR}~\cite{Routing1}. In the case of an $N \times N$ \ac{AWGR}, each source uses $N$ bits to encode which of its $N$ local wavelengths are occupied with one-hot encoding;
Even if we piggyback this information multiple times a second, the bandwidth impact is negligible.
For instance, if we multiplex multiple flows into a wavelength and therefore denote 8 bits per wavelength, the status vector per source becomes just $256 \times 8 = 2048 bits = 256 bytes$.
If, due to stale information, sources pick an intermediate destination whose direct wavelength to the final destination is not available, the intermediate destination performs indirect routing through a second intermediate destination. 
If no messages would otherwise be exchanged between a pair, thus presenting no opportunity for piggybacking, that pair can exchange a separate control message.

\subsection{Spatial and Wave-Selective Switches}
\label{section:spatial and wave selective switches}

Spatial and wave-selective switches can use indirect routing in tandem with reconfiguration. Indirect routing reduces the need for reconfiguration, but intermediate destinations should be chosen among destinations that already have a direct connection with the final destination; otherwise, the intermediate destination itself may trigger a reconfiguration. The synergy between indirect routing and switch reconfiguration was explored in~\cite{TAGO}. Our proposed design based on \acp{AWGR} avoids reconfiguration entirely (Section~\ref{section:available bandwidth}).

%% file: 5_system.tex
\section{Disaggregated Rack Design}
\label{section:disaggregated rack design}

For the rest of our study, we will model an \ac{HPC} rack based on a \ac{GPU}-accelerated HPE/Cray EX Supercomputer~\cite{Perlmutter_configuration}
where a rack contains 128 \ac{GPU}-accelerated nodes. Each node of our model system contains an AMD Milan \ac{CPU} with eight memory controllers, each supporting a 3200MHz DDR4 module. Therefore, each \ac{CPU} has 256 GB of memory with a maximum bandwidth of 204.8 GBps.
A compute node also has four NVIDIA Ampere A100 \acp{GPU}. Each \ac{GPU} supports 12 third generation NVLink links, each supporting 25 GBps per direction. Each \ac{GPU} also has 40 GB of co-located \ac{HBM} with a bandwidth of 1555.2 GBps. Each node also has four 31.5 GBps PCI Gen4 links to connect each \ac{GPU} to the \ac{CPU}. The \ac{CPU} also connects to four Slingshot 11 \acp{NIC} with 200 Gbps per direction~\cite{Slingshot_analysis}.

\subsection{MCMs and Escape Bandwidth}
\label{section:mcms and escape bandwidth}

We organize chips within each rack into an \acp{MCM} package. For simplicity, we restrict all \acp{MCM} to have the same escape bandwidth and we place chips of only the same type in \acp{MCM}. We then make conservative assumptions for next generation photonics that are entering the market today based on our analysis of Section~\ref{section:optical technologies for resource disaggregation}. In particular, each \ac{MCM} has 32 optical fibers attached to it, a conservative assumption compared to the five arrays of 24 fibers demonstrated in~\cite{Fibers_copackaged}.
Each fiber supports 64 wavelengths (channels) of 25 Gbps each for a 6400 GBps escape bandwidth per \ac{MCM}. We vary the number of chips per \ac{MCM} such that each chip enjoys the same escape bandwidth as in our baseline rack~\cite{Perlmutter_configuration}.
Therefore, our photonic architecture \emph{does not restrict chip escape bandwidth}. Table~\ref{table:MCM_types} shows the number of chips per \ac{MCM} and the total number of \acp{MCM} containing chips of that type to satisfy chip escape bandwidth. Each \ac{MCM} contains a controller chip that interfaces the native protocol of the disaggregated resource to the \ac{CXL} protocol over the photonic links. \ac{CXL}'s overhead and its associated \ac{FEC} is included in our architecture model.

\begin{table}
    \begin{center}
        \begin{tabular}{|l|c|c|}
            \hline 
            Chip type  & Chips per \ac{MCM} & \acp{MCM} per rack \\
            \hline
            \hline
            \ac{CPU}   &  14                 & 10                    \\
            \hline
            \ac{GPU}   &  3                  & 171                   \\
            \hline
            \ac{NIC}   &  203                & 3                     \\
            \hline
            \ac{HBM}  &   4                  & 128                   \\
            \hline
            DDR4   & 27                 & 38                    \\
            \hline
            \hline
            Total         &                    & 350                   \\
            \hline
        \end{tabular}
    \end{center}
    \caption{The number of chips of each type (\ac{CPU}, \ac{GPU}, \ac{NIC}, \ac{HBM}, or DDR4 module) per \ac{MCM} and \acp{MCM} in a rack assuming 32 fibers per \ac{MCM}, 64 wavelengths of 25 Giga bits per second per fiber. The target \ac{BER} to and from memory is $10^{-18}$ (Section~\ref{section:memorytechnology}).\label{table:MCM_types}}
    \ifnum\value{usevspace}>0
    \vspace{-1.0cm}
    \fi
\end{table}

\subsection{Optical Switches}
\label{section:optical switches}

The radix and wavelengths per port of optical switches dictate the number of \acp{MCM} we can fully connect optically with a single photonic switch as well as the amount of direct (single-hop) bandwidth. From Section~\ref{section:switch technologies}, we pick state-of-the-art representatives of wave-selective, cascaded \acp{AWGR}, and spatial optical switches. Their parameters are shown in Table~\ref{table:optical switch parameters}. Even though spatial~\cite{240x240WaferScale} and wave-selective switches~\cite{Hybrid_switch} are capable of 100 Gbps per wavelength, most links available widely today do not support that (Table~\ref{table:photonic link summary}). In addition, we show that we can still satisfy bandwidth demands with the conservative assumption of 25 Gbps per wavelength.

\begin{table}
    \begin{center}
        \begin{tabular}{|l|l|c|}
            \hline 
                                                  & Switch type                         & State of the art \\
            \hline
            \hline
            \multirow{3}{*}{Switch radix}         & Cascaded AWGRs~\cite{satoRealizationApplicationLargeScale2018}    & 370              \\
                                                  & Spatial~\cite{240x240WaferScale}    & 240              \\
                                                  & Wave-Selective~\cite{Hybrid_switch} & 256              \\
            \hline
            Gbps per wavelength                   & All switches                        & 25               \\
            \hline
            \multirow{3}{*}{Wavelengths per port} & Cascaded AWGRs~\cite{satoRealizationApplicationLargeScale2018}    & 370              \\
                                                  & Spatial~\cite{240x240WaferScale}    & 240              \\
                                                  & Wave-Selective~\cite{Hybrid_switch} & 256              \\
            \hline
        \end{tabular}
    \end{center}
    \caption{Switch configurations for our study.\label{table:optical switch parameters}}
    \ifnum\value{usevspace}>0
    \vspace{-0.8cm}
    \fi
\end{table}

\begin{figure}
  \centering
  \includegraphics[width=\columnwidth]{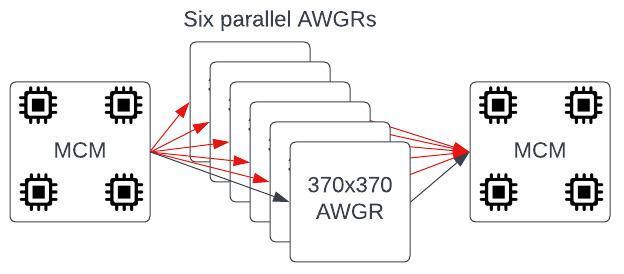}
  \ifnum\value{usevspace}>0
  \vspace{-0.6cm}
  \fi
  \caption{There are six parallel \acp{AWGR} that each \ac{MCM} connects to. There are at least five wavelengths (shown in red) between any particular \ac{MCM} pair.\label{figure:awgr case}}
  \ifnum\value{usevspace}>0
  \vspace{-0.6cm}
  \fi
\end{figure}

To connect our 350 \acp{MCM} using $370 \times 370$ \acp{AWGR}, we can combine \ac{MCM} fibers in five groups of six and connect each group to one port of five parallel \acp{AWGR}. However, each \ac{AWGR} port would be required to handle 384 wavelengths. To respect the per port 370 wavelength limitation of our \ac{AWGR} configuration but still satisfy the full escape bandwidth of \acp{MCM}, we combine the remaining 14 wavelengths along with the remaining two fibers per \ac{MCM} ($128 + 14 = 142$ wavelengths total) that were left unconnected into an extra parallel \ac{AWGR}, for a total of six parallel \acp{AWGR}. We then connect \ac{MCM} fibers to \acp{AWGR} in a staggered manner such that each \ac{MCM} connects to each other \ac{MCM} using at least five 25 Gbps direct-path wavelengths, for a direct \ac{MCM}--\ac{MCM} bandwidth of $25 \times 5 = 125$ Gbps. This is illustrated in \figurename~\ref{figure:awgr case}.

For simplicity, because of their relative small difference and because wave-selective switches can also achieve configurations that spatial switches can, we treat both wave-selective and spatial switches as 256 ports with 256 wavelengths per port. Each \ac{MCM} can connect to $\frac{2048}{256} = 8$ parallel switches. However, because the radix of optical switches is lower than the number of \acp{MCM}, we instantiate 11 optical switches and connect \acp{MCM} in a staggered manner such that an optical switch with an index $I$ connects to \acp{MCM} that have an index starting from $(32 \times I) \mod 350$ until $(I + 255) \mod 350$. This way, a small number of optical switch ports are left unconnected to not exceed the 32 fibers per \ac{MCM}. These ports can support larger racks in the future. If the switches configure appropriately, each \ac{MCM} has at least three direct paths to any other \ac{MCM}. Each path has 256 wavelengths, thus, the direct \ac{MCM} bandwidth is $256 \times 3 \times 25 = 2304$ Gbps.

%% file: 6_evaluation.tex
\section{Evaluation}
\label{section:evaluation}

Having evaluated in Section~\ref{section:bit error rates and fec} that photonic switches satisfy \ac{BER} requirements, we now analyze the impact of photonic-based intra-rack resource disaggregation on bandwidth, latency, and power.

\subsection{Bandwidth Evaluation}
\label{section:photonic switch comparison}

We distinguish two cases based on Section~\ref{section:optical switches}: (A) Six parallel \acp{AWGR} and (B) 11 parallel wave-selective switches.

\subsubsection{Available Bandwidth}
\label{section:available bandwidth}

Using either or both indirect routing and switch reconfiguration, any one particular \ac{MCM} can use its full escape bandwidth to reach a single destination \ac{MCM}. In case (A), all wavelengths escaping an \ac{MCM} can reach the same destination \ac{MCM} using only indirect routing (since \acp{AWGR} do not reconfigure). In case (B), the photonic switch itself can reconfigure to route $768$ wavelengths directly to a destination \ac{MCM}; the other $2048 - 768 = 1280$ wavelengths can be configured to route indirectly through intermediate \acp{MCM}. Therefore, while both cases (A) and (B) can provide the same source--destination bandwidth assuming no contention, spatial and wave-selective switches have to use a centralized scheduler that is prone to making imperfect decisions and imposes power and latency overheads~\cite{Switch_scheduler}. In contrast, case (A) only uses distributed indirect routing that avoids much of those overheads.

Based on profiling data of a production open-science \ac{HPC} system~\cite{TACO_disaggregation}, the 125 Gbps direct bandwidth between \acp{MCM} in case (A) suffices over 99.5\% of the time between \acp{CPU} and main memory (DDR4) and virtually all the time between memory and \acp{NIC}. In addition, the bandwidth of a single \ac{AWGR} wavelength of 25 Gbps suffices 97\% of the time between \acp{CPU} and memory as well as between memory and \acp{NIC}. This means that with a 97\% probability, four of the five wavelengths between memory--\ac{CPU} or \ac{NIC}--memory pairs are available to use for indirect routing in case the direct 125 Gbps bandwidth does not suffice between another memory--\ac{CPU} or \ac{NIC}--memory pair. Therefore, the probability at any one time that the direct bandwidth does not suffice for a number of \ac{CPU}--memory and \ac{NIC}--memory pairs large enough such that they cannot find unused bandwidth in other pairs to use with indirect routing is negligible. To further reduce the probability, congested pairs can use direct paths from \acp{CPU} to other \acp{CPU} that communicate minimally and \acp{NIC} to other \acp{NIC} that typically do not communicate~\cite{TACO_disaggregation}.
Thus, case (A) satisfies bandwidth between \acp{CPU}, \acp{NIC}, and main memory (DDR4).

\begin{figure*}[ht]
  \begin{center}
    \subfigure{%
      \includegraphics[width=0.49\textwidth]{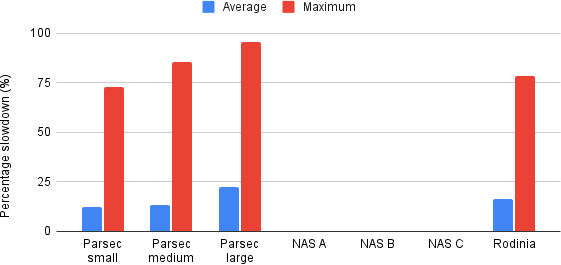}
      \label{subfig:inorder cpu slowdown}
    }
    \hfill
    \subfigure{%
      \includegraphics[width=0.49\textwidth]{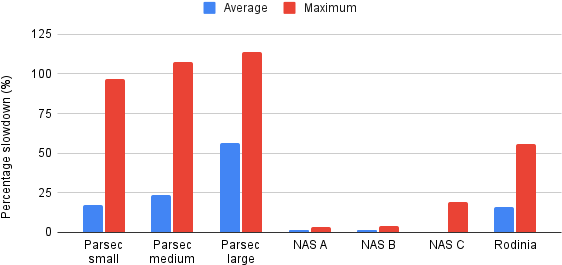}
      \label{subfig:ooo cpu slowdown}
    }
    \ifnum\value{usevspace}>0
    \vspace{-0.8cm}
    \fi
    \caption{Average and maximum slowdown for each benchmark suite and input set size. The slowdown is for an additional 35ns of latency between the \ac{LLC} and main memory from the additional photonic components. Left: in-order pipeline compute cores. Right: Out of order (OOO) compute cores.\label{figure:cpu slowdown}}
    \ifnum\value{usevspace}>0
    \vspace{-0.8cm}
    \fi
  \end{center}
\end{figure*}

For \acp{GPU}, in case (A) with indirect routing, a single \ac{GPU} can use a total of $125 \times 512 = 8000$ GBps to access any one \ac{HBM} or more in case a \ac{GPU} is allocated more than one \acp{HBM}. This well satisfies the 1555.2 GBps that NVIDIA Ampere A100 \acp{GPU} in our model rack~\cite{Perlmutter_configuration} access \ac{HBM} with today, and leaves $8000 - 1555.2 = 6444.8$ GBps unused per \ac{GPU}. In addition, in the worst case, an \ac{MCM} containing three \acp{GPU} will communicate at full bandwidth (12 NVLink links of 25 GBps per each of the three \ac{GPU} equals 900 GBps) to other \acp{MCM} containing \acp{GPU}. Here, if all \acp{GPU} in the rack act similarly, we cannot rely on indirect routing from a \ac{GPU} through an intermediate \ac{GPU} to reach a destination \ac{GPU}. The direct 125 Gbps bandwidth between \ac{GPU} \acp{MCM} does not suffice. Therefore, each \ac{GPU} can use the 6444.8 GBps of unused bandwidth to and from \acp{HBM} for indirect routing to sufficiently cover the 900 GBps bandwidth that would otherwise use NVLink \ac{GPU}--\ac{GPU} links. This leaves $6444.8 - 900 = 5544.8$ GBps per \ac{GPU} that can support direct \ac{HBM}--\ac{HBM} communication such as due to GPUDirect RDMA, indirect routing for other \acp{MCM}, or simply increase available bandwidth to memory. Notably, our analysis does not use direct optical paths from \acp{GPU} to main memory (DDR4). Future protocols may use these paths, or they can provide even more indirect routing bandwidth.

Our analysis shows that case (A) with \acp{AWGR} more than satisfies bandwidth demands and avoids the need for a scheduler to reconfigure spatial and wave-selective switches that would otherwise add overhead and increase reaction time.

\subsection{Latency Evaluation}
\label{section:impact of photonic latency to memory}

For intra-rack disaggregation, we assume an additional latency between \acp{MCM} of 35 ns, significantly less than full system disaggregation. That additional latency covers 15 ns for electrical--optical--electrical conversion and 4 meters of photonic propagation at 5 ns per meter, which covers the round-trip distance of typical two-meter tall racks (Section~\ref{section:link propagation and encoding/decoding latency}). The small impact of distance to latency with photonics practically makes \acp{MCM} within a rack equidistant; this mitigates a traditional queuing delay versus locality tradeoff in job scheduling~\cite{GPU_profiling_2019} and reduces the need for hierarchical memory~\cite{Disaggregation_tiered_memory,Disaggregation_page_policy}.
Indirect routing would increase latency by a few extra ns, but the probability of routing indirectly is low. Also, because 35 ns is orders of magnitude lower than system-wide network latency, we do not consider the effect of the additional 35 ns on inter-rack communication (such as traditional MPI) through \acp{NIC}.

\subsubsection{CPU Evaluation}
\label{section:cpu evaluation}

We experimentally quantify the impact of the additional latency on application performance with in-order pipelined and \ac{OOO} compute cores. In-order cores provide clear insight into the impact of memory latency because in-order cores do not mask latency, whereas \ac{OOO} cores are representative of modern systems. We use full system simulation in Gem5~\cite{Gem5}
of x86 compute cores running an Ubuntu 18.4 guest \ac{OS}. We configure the cache hierarchy to match the \acp{CPU} of our model \ac{HPC} rack~\cite{Perlmutter_configuration}. We calculate the slowdown of application execution time when we add 35 ns of latency between the \ac{LLC} and main memory, compared to a baseline system with no additional latency to memory.
Latency is the only potential source of application slowdown since our architecture satisfies the full escape bandwidth of each chip.

We evaluate the impact to three benchmark suites: PARSEC 3.1~\cite{Parsec}, NAS parallel benchmarks 3.4.1~\cite{NAS}, and Rodinia~\cite{Rodinia}. For PARSEC, we evaluate small, medium, and large input sets. For NAS, we evaluate input sizes ``A'', ``B'', and ``C''. For Rodinia, we use the single default input set. These benchmark suites have been widely used and contain a large variety of computation kernels that are representative of key \ac{HPC} applications such as stencils, graph processing, linear algebra, computational mathematics, grid, sorting, and many others that have been observed to be important workloads in NERSC systems~\cite{N10_workload_analysis}.
Overall, we use 57 \ac{CPU} benchmarks for a broad representation.
We use a single compute core to better focus on the effect of the additional latency to memory.

\figurename~\ref{figure:cpu slowdown} shows slowdown percentages for benchmarks across our three suites for an in-order core on the left and an \ac{OOO} core on the right. As shown, NAS benchmarks are negligibly affected by the increased latency from photonics. Rodinia benchmarks have an average slowdown of 16\% for both in-order and \ac{OOO} cores. Benchmark NW shows the largest slowdown of approximately 79\% for in-order cores and 55\% for \ac{OOO} cores. For Parsec benchmarks with large inputs, the average slowdown is 23\% for in-order cores and 41\% for \ac{OOO} cores. However, for medium inputs, those slowdowns drop to 13\% and 24\%, respectively, because with medium inputs more benchmarks have a working set that fits in the \ac{LLC}. The overall average slowdown for Parsec across input sizes is 16\% for in-order cores and 27\% for \ac{OOO} cores. Across all benchmarks of the three suites and input sizes, the average slowdown with in-order cores is 15\% and with \ac{OOO} cores 22\%.
These slowdowns are considerably less than the slowdowns quoted in past work for full-system disaggregation (Section~\ref{section:related work}).

\begin{figure*}[ht]
  \begin{center}
    \subfigure{%
      \includegraphics[width=0.49\textwidth]{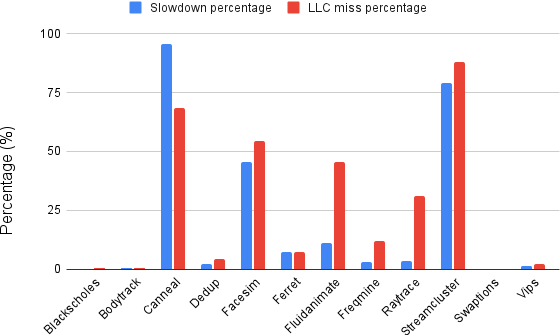}
      \label{subfig:parsec caches}
    }
    \hfill
    \subfigure{%
      \includegraphics[width=0.49\textwidth]{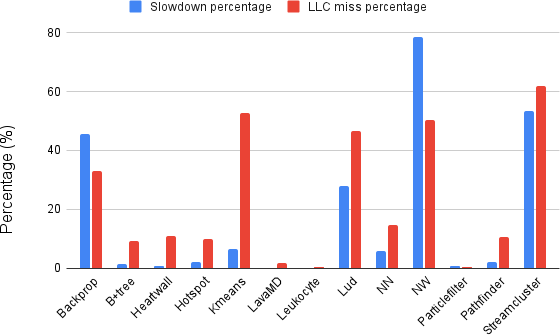}
      \label{subfig:rodinia caches}
    }
    \ifnum\value{usevspace}>0
    \vspace{-0.8cm}
    \fi
    \caption{Slowdown of Parsec benchmarks with large inputs (left) and Rodinia benchmarks (right) for in-order cores. In addition, each benchmark's average \ac{LLC} miss rate is shown. Benchmarks with larger miss rates produce higher slowdowns.\label{figure:cache graphs}}
    \ifnum\value{usevspace}>0
    \vspace{-0.8cm}
    \fi
  \end{center}
\end{figure*}

\figurename~\ref{figure:cache graphs} illustrates that the performance penalty correlates with the \ac{LLC} miss rate. In fact, for Parsec with large inputs, the Pearson product-moment correlation coefficient is 0.89, while for Rodinia 0.76. For all Parsec benchmarks with small, medium, and large inputs, the coefficient is 0.822. While not shown, \ac{OOO} cores show a similar behavior partly because they do not substantially change the \ac{LLC} access patterns or working set sizes. In fact, for Rodinia with \ac{OOO} cores, the correlation factor is 0.93, while for Parsec 0.75.
All these coefficients indicate a strong correlation.
In addition to \ac{LLC} miss rate, the ratio of memory accesses to non-memory instructions is a key factor to an application's slowdown;
\ac{OOO} cores de-emphasize this parameter slightly and stress the \ac{LLC} miss rate more, as mentioned above.

Furthermore, we notice that the cycles the \ac{LLC} spends in a miss
increase by 50\% to 150\% across benchmarks for in-order and \ac{OOO} cores. We further confirm the importance of \ac{LLC} miss rates to performance by observing that streamcluster with small and medium inputs has an \ac{LLC} miss rate of less than 0.5\% and a negligible slowdown. However, streamcluster with large inputs has a working set that does not fit in the \ac{LLC}, causing an \ac{LLC} miss rate of over 60\% and thus a slowdown of about 57\%.

Focusing on the performance of individual benchmarks, for in-order cores, only three benchmarks exceed a 25\% slowdown in each of Rodinia and Parsec (large) whereas for \ac{OOO} cores only two benchmarks in Rodinia and three in Parsec (large).
Therefore, the majority of benchmarks are impacted lightly, even without mitigation strategies. For more affected benchmarks, there is a range of mitigating hardware and software techniques~\cite{Latency_tolerance, Latency_tolerance2, Latency_tolerance3, Latency_tolerance4}.
Our results motivate more memory latency-tolerant compute units for resource-disaggregated systems.

\begin{figure*}[ht]
  \begin{center}
    \subfigure{%
      \includegraphics[width=0.49\textwidth]{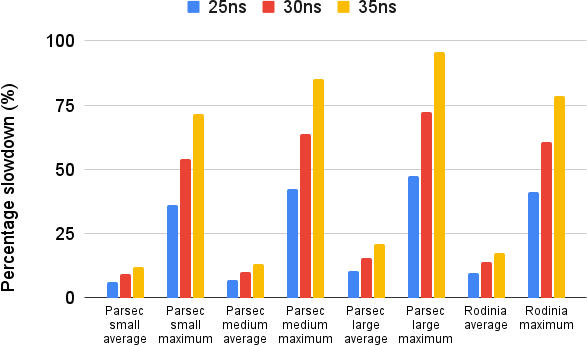}
      \label{subfig:varying latencies inorder}
    }
    \hfill
    \subfigure{%
      \includegraphics[width=0.49\textwidth]{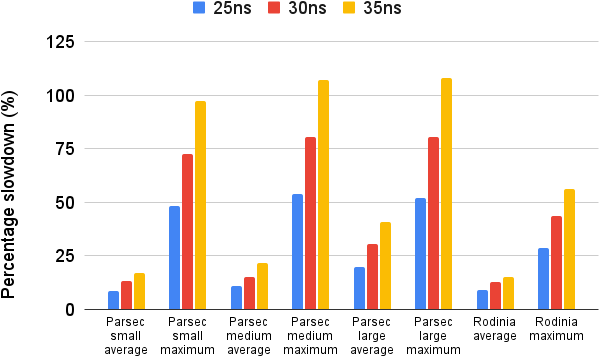}
      \label{subfig:varying latencies ooo}
    }
    \ifnum\value{usevspace}>0
    \vspace{-0.6cm}
    \fi
    \caption{Slowdown for 25 ns, 30 ns, 35 ns of additional \ac{LLC}-memory latency for in order (left) and \ac{OOO} cores (right).\label{figure:varying latencies}}
    \ifnum\value{usevspace}>0
    \vspace{-0.8cm}
    \fi
  \end{center}
\end{figure*}

\subsubsection{Sensitivity to Latency}
\label{section:sensitivity to latency}

Thus far, we assumed 35 ns to cover 4 meters, which is the worst-case intra-rack distance in modern systems. Here, we assess whether improved photonics or shorter rack distances with lower latencies would greatly benefit application performance by comparing performance for 25 ns, 30 ns, and 35 ns. Results are shown in \figurename~\ref{figure:varying latencies}. For both in-order and \ac{OOO} cores, reducing the additional latency to 25 ns from 35 ns reduces application slowdown by about half.

\subsubsection{GPU Evaluation}
\label{section:gpu evaluation}

To evaluate the impact of the additional latency between \acp{GPU} and \ac{HBM},
we extend the publicly available version of PPT-GPU~\cite{PPT-GPU-SC21} toolkit to account for the additional latency between the main memory of the \ac{GPU} and the \ac{LLC}. In our evaluation, we model one NVIDIA A100 \ac{GPU}~\cite{A100} running a total of 24 applications with 1525 kernels from different benchmark suites. We run 11 applications from Rodinia~\cite{Rodinia} and ten applications from Polybench~\cite{polybench}. Polybench applications are linear algebra applications that stress the \ac{GPU} cache and main memory. Furthermore, we run AlexNet, GRU, and LSTM from the Tango deep network~\cite{tango} benchmark suite. We use the default input sizes and configurations that came with the benchmarks, detailed in~\cite{PPT-GPU-SC21}. We run applications using the ``SASS'' model and extract memory and instruction traces for each application.

\begin{figure}
  \centering
  \includegraphics[width=0.9\columnwidth]{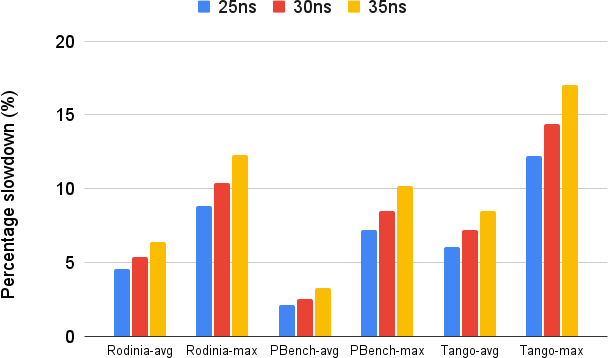}
  \ifnum\value{usevspace}>0
  \vspace{-0.4cm}
  \fi
  \caption{Slowdown for 25 ns, 30 ns, and 35 ns of additional \ac{LLC}--memory latency for different \ac{GPU} benchmarks.\label{figure:GPGPU-results}}
  \ifnum\value{usevspace}>0
  \vspace{-0.6cm}
  \fi
\end{figure}

\begin{figure}
  \centering
  \includegraphics[width=\columnwidth]{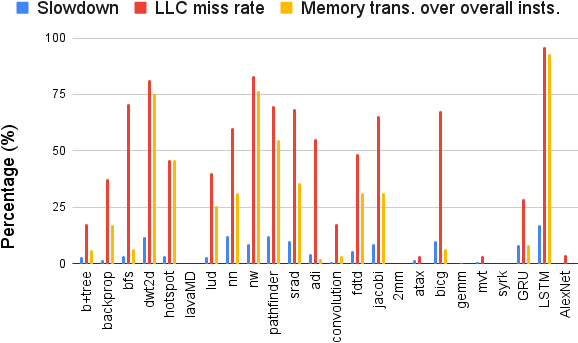}
  \ifnum\value{usevspace}>0
  \vspace{-0.8cm}
  \fi
  \caption{Slowdown for 35 ns, \ac{LLC} miss rate, and memory (\ac{HBM}) transactions over total instructions per \ac{GPU} benchmark.\label{figure:GPGPU-caches}}
  \ifnum\value{usevspace}>0
  \vspace{-0.8cm}
  \fi
\end{figure}

\figurename~\ref{figure:GPGPU-results} shows the effect of different latencies on the performance of our \ac{GPU} benchmarks. We compare performance in terms of the total predicted cycles. The average slowdown across all 24 \ac{GPU} applications is 5.35\%. In addition, \figurename~\ref{figure:GPGPU-caches} shows that the slowdown (shown for 35 ns) has a strong correlation with (i) the \ac{LLC} miss rate and (ii) the percentage of transactions the \ac{HBM} receives over the total number of instructions, indicated by a correlation factor of 0.87 and 0.79, respectively. In contrast, it has no significant correlation with the percentage of memory request instructions over the total number of instructions because the caches filter a different percentage of those requests.

\subsubsection{CPU--GPU Comparison}
\label{section:cpu gpu comparison}

\begin{figure}
  \centering
  \includegraphics[width=0.9\columnwidth]{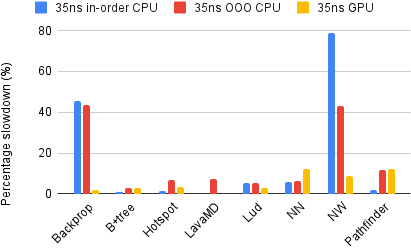}
  \ifnum\value{usevspace}>0
  \vspace{-0.4cm}
  \fi
  \caption{Slowdown for \ac{CPU} and \ac{GPU} Rodinia benchmarks.\label{figure:cpu_gpu}}
  \ifnum\value{usevspace}>0
  \vspace{-0.6cm}
  \fi
\end{figure}

We illustrate the difference in memory latency tolerance of in-order \acp{CPU}, \ac{OOO} \acp{CPU}, and \acp{GPU} in \figurename~\ref{figure:cpu_gpu} for the intersection of Rodinia benchmarks that correctly complete on both \acp{CPU} and \acp{GPU} with their default input sets. As shown, \acp{GPU} tolerate the additional 35 ns latency better with a maximum slowdown of 12\%. This is promising for resource disaggregation given the steady growth of \acp{GPU} in \ac{HPC} systems.

\subsection{Power Overhead}
\label{section:power overhead}

We calculate the per-rack power overhead of our photonic solution for 350 \acp{MCM} with 2048 escape wavelengths from each \ac{MCM} and 25 Gbps per wavelength. If we use demonstrated comb laser transceiver pairs that consume approximately 0.5 pJ/bit including laser power~\cite{New_energy_links1,Energy_efficient_links2}, and include the switches of Table~\ref{table:photonic switch summary} that consume in total for all parallel switches no more than 1 kW, the total additional power for all photonic components is approximately 11 kW. Our analysis pessimistically assumes photonic components are constantly on. Considering that the power consumption of an A100 \ac{GPU} is approximately 300 W, an AMD Milan \ac{CPU} 250 W, and 512 GB of DDR4 memory in a single node approximately 192 W, the power overhead for our photonic solution is approximately 5\%.

\subsection{Comparison With Electronic Switches}
\label{section:comparison with electronic switches}

The electronic SERDES signaling rate per wire is only 112 Gbps for a short reach. Also, typical \ac{CXL} or PCIe signaling rates top out at 35 GHz/wire. In fact, as SERDES rates increase, the distance that those signals can reach reduces to even a few millimeters due to the resistance and capacitance of copper wires. Photonics break the reach limitations of copper and, with co-packaging, can achieve 4 Tbps per mm of shoreline on the chip die.

Focusing on electronic switches, Rosetta~\cite{Slingshot_latency} and Infiniband~\cite{Infiniband_latency} have a measured per hop latency of no less than approximately 200 ns. Emerging PCIe Gen5 switches add just 10 ns per hop~\cite{PCIe_gen5}, but only support 100 lanes per switch. To fully connect our disaggregated rack, we consider a two-level tree network with four hops (the top level is composed of an internal two-hop subnetwork). These four hops will be in addition to the 35 ns we previously evaluated for \ac{FEC} and propagation (propagation delay is comparable between copper and photonic for intra-rack distances), since our photonic solution uses switches with negligible traversal latency. Therefore, the additional latency for disaggregation in the PCIe case becomes 85 ns compared to 35 ns for our photonic architecture. Finally, we also consider the latency through one hop of an Anton 3 network, which is approximately 90 ns by average~\cite{Anton3}, though scaling up to match our rack size would require multiple hops.
These latencies are optimistic thus favorable for electronic switches considering that recent small-group prototypes using \ac{CXL} report a minimum of 142 ns latency~\cite{Disaggregation_pond}. Scheduler decisions or congestion can cause higher worst-case (tail) latencies that may further penalize application performance.
This assumes that we connect only one lane per endpoint, which carries 32 Gbps for PCIe Gen5 and 29 Gbps for Anton 3.
This is multiple times less than the per-chip bandwidth of our photonic architecture.

\begin{figure}
  \centering
  \includegraphics[width=\columnwidth]{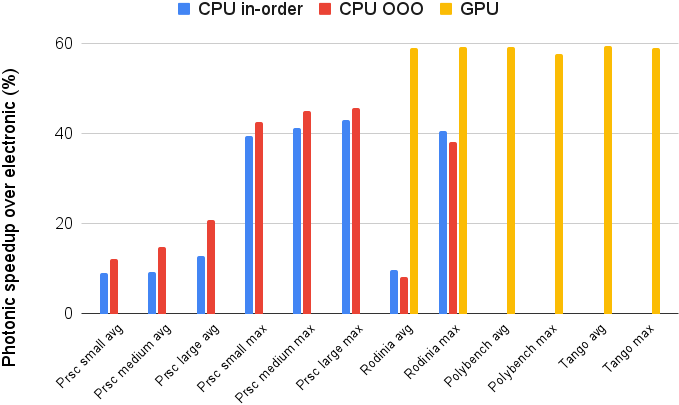}
  \ifnum\value{usevspace}>0
  \vspace{-0.8cm}
  \fi
  \caption{Speedup of a system that uses emerging photonics to implement intra-rack resource disaggregation that adds 35 ns of additional latency to and from memory compared to a similar system that uses modern electronic switches and adds 85 ns of memory latency instead.\label{figure:electronic_switch_slowdown}}
  \ifnum\value{usevspace}>0
  \vspace{-0.6cm}
  \fi
\end{figure}

\figurename~\ref{figure:electronic_switch_slowdown} shows the speedup of a system that implements intra-rack disaggregation with emerging photonics with an additional 35 ns latency to and from DDR4 and \ac{HBM} memory compared to a similar system that uses modern electronic switches instead. 85 ns is currently the lowest latency for electronic switches and corresponds to a four-hop PCIe Gen5 network or a single-hop Anton 3 network.
As shown, for \ac{CPU} benchmarks, if we only take into account ``medium'' from PARSEC to avoid counting PARSEC benchmarks three times, the average speedup for in-order cores is 9\% and the maximum 41\%. For \ac{OOO} compute cores, the average is 15\% and the maximum 45\%.
We notice that electronic switches increase the \ac{LLC}'s total miss cycles by approximately 100\% to 150\%. For \acp{GPU}, the average and maximum are both 61\%.
These results show that the reduced latency of photonics compared to electronic switches has a significant application impact, making disaggregation with photonics more attractive.

\subsection{Iso-Performance Comparison}
\label{section:iso-peformance comparison}

Based on our performance evaluations, we estimate that in order to preserve system-wide average computational throughput as our baseline \acs{GPU}-accelerated HPE/Cray EX system~\cite{Perlmutter_configuration}, our photonically-disaggregated system requires 6\% more \acp{GPU} and 15\% more \acp{CPU}, assuming in-order \acp{CPU} which is the worst case. However, intra-rack resource disaggregation allows our rack to have an average 4$\times$ fewer memory modules and 2$\times$ fewer \acp{NIC}~\cite{TACO_disaggregation}. Combining the two effects, our disaggregated rack has
1075 total modules compared to 1920 in the equal-performance baseline system, an approximately 44\% reduction. With such a reduction, the reduced overhead for power distribution and cooling more than compensates for the negligible power increase from photonics. Alternatively, we can preserve all rack resources and instead add 128 of a combination of \acp{CPU} and \acp{GPU} (with their \acp{HBM}), which is only an approximately 7\% chip increase compared to a rack of the baseline system. Doing so \emph{doubles} computational throughput.

%% file: 7_discussion.tex
\section{Discussion}
\label{section:discussion}

Our study highlights that more latency-tolerant \acp{CPU}~\cite{Memory_latency_tolerance} and \acp{GPU}~\cite{GPU_memory_tolerance} would make resource disaggregation more attractive. This insight is also important for systems with compute accelerators and \acp{FPGA}. \acp{FPGA} can better tolerate memory latency by customizing their compute logic~\cite{FPGA_latency_tolerance}, prefetching~\cite{FPGA_prefetching}, multithreading~\cite{FPGA_multithreaded}, and burst scheduling~\cite{FPGA_bursts}.
Accelerators can use customized prefetching~\cite{Accelerator_prefetching, Accelerator_prefetching2} and other techniques~\cite{Accelerator_latency_incensitive,CNN_latency_hiding,Memory_latency_sweep}.

While Perlmutter is a top \ac{HPC} system, other systems should repeat our analysis to design their disaggregation hardware. Chips with higher escape bandwidths motivate fewer chips per \ac{MCM} and more parallel \acp{AWGR}, but do not increase chip-to-chip photonic latency. The diversity of bottlenecks in \ac{HPC} applications is a motivating argument for resource disaggregation that provides the ability to change the balance of node resources and thus support the diversity of scenarios that are typically present in a mixed-workload system.

%% file: 8_conclusion.tex
\section{Conclusion}
\label{section:conclusion}

We discuss a resource-disaggregated \ac{HPC} rack that uses modern photonic links and switches to meet \ac{BER} and bandwidth requirements of \ac{HPC} applications, has just a 5\% power overhead, uses distributed indirect routing, is faster than an equivalent architecture with electronic switches, and allows an iso-performance system to have 44\% fewer chips.

\section*{Acknowledgments}
This work was supported by ARPA-E ENLITENED Program (project award DE-AR00000843) and the Director, Office of Science, of the U.S. Department of Energy under Contract No. DE-AC02-05CH11231.